ability, but also for its use of citizen science as an emerging data source. The project explores the spectrum of citizen science ranging from least to most buy-in: from initial engagement (interactive awareness raising) to crowdsourcing (collecting data from the public) to participatory action (citizens collect data and analyse it). The three citizen engagement tools used in the project are:

1. The Individual Weather Map [L1] is a way that citizens can engage on an individualised level with their local urban climate conditions using an interactive slider to indicate levels of personal temperature preference. Through this, citizens will be able to interact with weather/climate information (e.g. planning their commute to pass through cooler areas), which will motivate and spark interest in climate action.
2. The Historical Weather Data Collection hub allows citizens to submit any personal weather records/observations in CSV format [L2] or through a survey form [L3]. Crowdsourcing historical weather data helps us gain insights into local weather and climate related processes, with a focus on providing knowledge about climate needed for the lives of citizens. It also shows citizens that climate change is already visible and detectable in their local community.
3. A major aspect of the project is collection of in situ climate data (using weather stations – see Figure 1, and mobile sensors for bicycles – see Figure 2), which can be used to validate other sensor systems in the project and ground-truth climate data [L4]. The data will provide opportunities for co-design methodologies for city climate adaptation where citizens and decisions-makers work together. One advantage of co-design is that the participatory role of citizens will more likely lead to policies and adaptation measures that are more relevant to the communities they serve.

The value proposition of CityCLIM is that no other service has explored urban climate using an operational weather model enriched with in situ and EO data while integrating citizen science. There is high potential that lies in the knowledge of local conditions and necessity of working with communities toward climate adaptation strategies. Facilitating citizen knowledge-sharing in decision-making can also foster a greater sense of civic duty and future climate action. Overall, citizen science is a more responsible and inclusive scientific methodology, where full-time and volunteer experts can learn from and with each other on an equal footing. These advantages are what the CityCLIM project aims to yield with its citizen science tools for urban climate monitoring.

**Links:**
[L1] https://www.rtl.lu/meteo/cityclim
[L2] https://meteologix.com/ua/info/citizenscience
[L3] https://survey.hifis.dkfz.de/544485/lang/en/
[L4] https://tinyurl.com/ymu7vv25

**Reference:**
[1] UNEP, "Cities and climate change," n.d. [Online]. Available: https://www.unep.org/explore-topics/resource-efficiency/what-we-do/cities/cities-and-climate-change.

**Please contact:**
Christine Liang, Helmholtz Centre for Environmental Research, Germany
christine.liang@ufz.de

# Temperature Monitoring of Agricultural Areas in a Secure Data Room

by Thomas Ederer, Martin Ivancsits and Igor Ivkić (FH Burgenland)

*Agricultural production is highly dependent on naturally occurring environmental conditions like change of seasons and the weather. Especially in fruit and wine growing, late frosts occurring shortly after the crops have sprouted have the potential to cause massive damage to plants [L1,L2] [1]. In this article we present a cost-efficient temperature monitoring system for detecting and reacting to late frosts to prevent crop failures. The proposed solution includes a data space where Internet of Things (IoT) devices can form a cyber-physical system (CPS) to interact with their nearby environment and securely exchange data. Based on this data, more accurate predictions can be made in the future using machine learning (ML), which will further contribute to minimising economic damage caused by crop failures.*

The production of food in agriculture often follows traditional practices, relying on the knowledge passed down from previous generations and on intuition-based decisions. These decisions relate to activities such as sowing, the application of fertilisers, the protection of crops and harvesting. In addition, agriculture is strongly influenced by natural conditions such as location, climate and weather. In sectors such as fruit and wine growing, frosts that occur shortly after crops have started to grow can cause significant damage. As a result, that year's crop is often significantly smaller and of lower quality, leaving the affected farms with some permanent damage. The knock-on effects extend to the market, where such late frosts can lead to scarcer, lower-quality produce and higher prices for consumers. In the most severe cases, some crops may not be available in the local area for an entire season. This requires substitutes to be sourced from distant locations, which increases emissions and has a negative impact on the climate.

Historically, frost damage has been mitigated using a variety of techniques, including water spraying, heaters and fumigation. These methods, while using existing weather systems, are largely manual and often overlook unique local features such as field alignment, wind shelters or nearby water sources. As a result, certain parts of the farm may experience dangerous drops in temperature that can go unnoticed by the system. In anticipation, some farmers employ people to monitor their fields on particularly cold nights, sounding the alarm when temperatures reach critically low levels. Although proactive, this method is expensive, physically demanding and increases the likelihood of human error. In addition, when applying mitigation measures, farmers tend to err on the side of caution and use resources such as water and fuel more liberally than is necessary.

To reduce manual effort in the field and save resources, we present an end-to-end use case of an International Data Space (IDS) [L3] designed for detailed temperature monitoring in agricultural regions [2,3]. The IDS consists of IoT devices

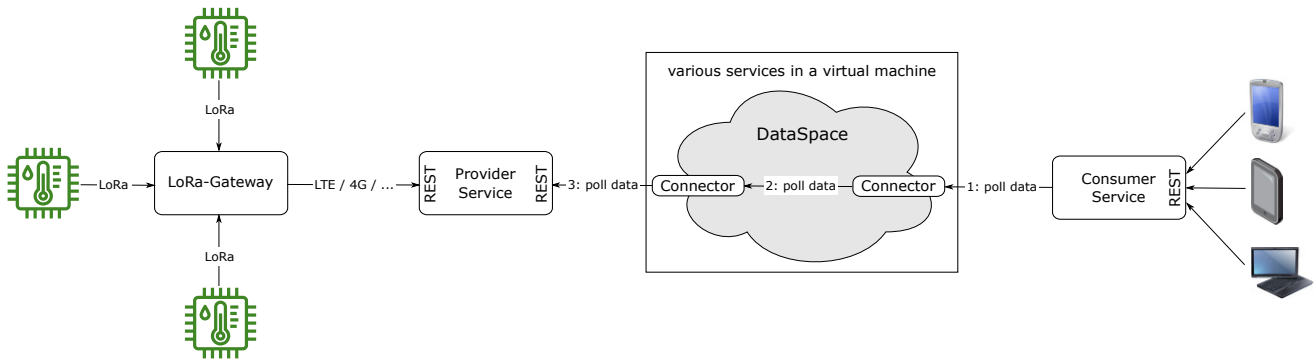

*Figure 1: Architecture of OrViCon.*

with temperature sensors that periodically transmit their data to a long-range (LoRa) gateway via the LoRa wide area network (LoRaWAN). The gateway then transmits this data via long-term evolution (LTE) to a provider service, which is managed by a cloud service provider (CSP). The provider service stores the sensor data and makes it available to other (requesting) IDS members. To ensure that the sensors-owning farm remains in full control of their data, the measurements are not automatically transmitted to other systems. Instead, within the IDS, the data provider only shares its data on request and only with a data consumer that is also part of the same data space. The following figure shows the architecture of the proposed IDS-based Orchard/Vinyard Control (OrViCon) temperature monitoring system:

As shown in Figure 1, the OrViCon architecture consists of temperature sensors on the edge, while the provider service and data space are running in the cloud. The temperature sensors installed in the agricultural area periodically send their data via LoRaWAN [L4] to the LoRa gateway, which then forwards the data to a provider service hosted by a CSP. The provider service adds the GPS coordinates of the corresponding sensor and stores the data for a potential requesting consumer service. The data space provides the necessary connectors to establish a secure end-to-end connection between a consuming service to a data-providing service. This approach guarantees that only data space members have access to certain datasets and that the data is only transmitted securely upon a request from a consumer service.

As shown in Figure 2, the IDS testbed consists of two connector instances: the data provider and the data consumer. The two connector instances are configured manually to specify which dataset can be offered by the data provider, or which dataset can be requested by a data consumer. The proposed OrViCon monitoring system provides a suitable solution for data exchange in agriculture scenarios. A key benefit of the system is that every new member must enrol prior to accessing other services within the data space. Another advantage is that, after successful registration, members are required to choose only from approved (certified) connectors to establish a connection to the data space.

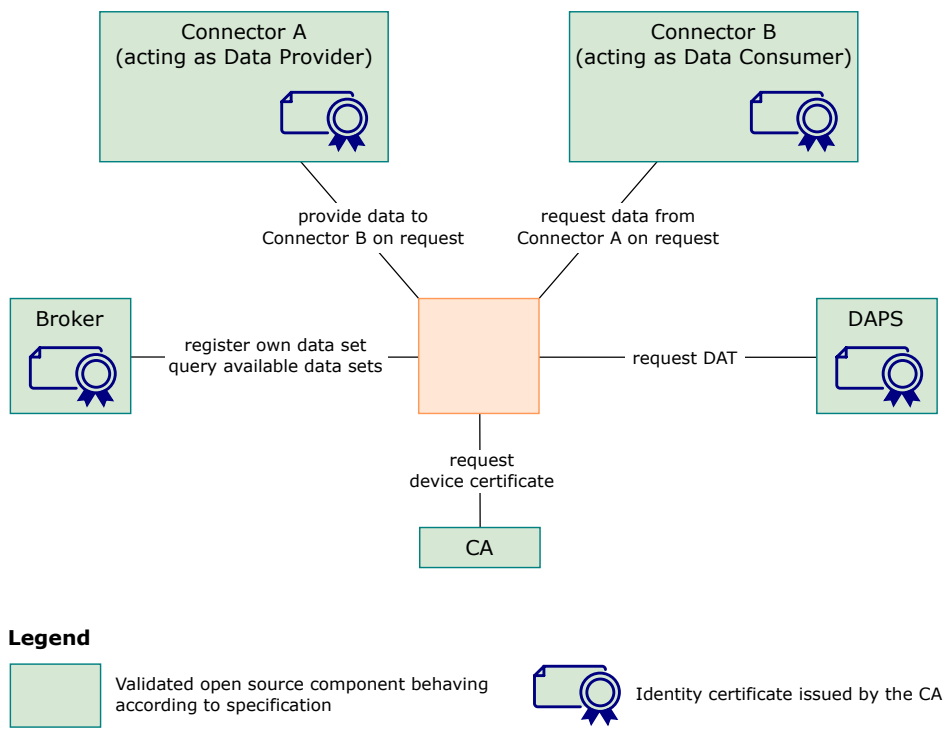

*Figure 2: Structure of the IDS testbed [L5].*

In conclusion, the data space provides the necessary software architecture to enable sovereign and secure data exchange between registered, trusted members. This ensures that the measurement data provided can only be accessed by certified members within the data space. In addition, data providers can set conditions (or rules) for the measurement data they make available. Potential consumers must agree to these conditions before they can use the data [L7]. The proposed OrViCon monitoring system shown in Figure 1 provides temperature data for an agricultural field that can be used to guide targeted measures against late frost. Based on the measured data, mitigation measures can be localised to specific areas rather than being applied across the entire agricultural landscape. This enables a more precise and economical use of resources such as water and fuel, resulting in reduced pollutant emissions and a positive impact on the climate.

Links:
[L1] https://tinyurl.com/5f6r7yt8
[L2] https://tinyurl.com/5dn5xtth
[L3] https://internationaldataspaces.org/
[L4] https://lora-alliance.org/
[L5] https://tinyurl.com/2b8rtsam
[L6] https://tinyurl.com/bdhxs5xx
[L7] https://tinyurl.com/2p9hf5k5

References:
[1] M. R. Salazar-Gutiérrez, B. Chaves, G. Hoogenboom, G., "Freezing tolerance of apple flower buds", Scientia Horticulturae, vol. 198, pp. 344–351, 2016.
[2] E. Curry and S. Scerri, T. Tuikka, "Data spaces: design, deployment and future directions", p. 357, Springer Nature, 2022.
[3] H. Ding, L. Liu, Z. Liang, "Research on environmental monitoring and prediction method of orchard frost based on wireless sensor network", in Int. Conf. on Guidance, Navigation and Control, pp. 7323–7332, Singapore, Springer Nature Singapore, 2022.

Please contact:
Thomas Ederer and Martin Ivancsits
FH Burgenland, Austria
2210781006@fh-burgenland.at,
2210781020@fh-burgenland.at

# A Mobile Phone App for Measuring Food Waste in Greek Households

by Prokopis K. Theodoridis, (Hellenic Open University), Theofanis V. Zacharatos and Vasiliki S. Boukouvala (University of Patras)

*A new mobile app has been developed in Greece to track household food waste. The app is the first of its kind in the country and has the potential to significantly reduce food waste. The app allows users to easily record how much food they waste each day. The data collected by the app can then be used to identify areas where food waste is most prevalent and to develop strategies for reducing it. The data processing reveals that an average Greek household wastes around 400 portions of food annually, with an economic cost of €800–1,000. The app has the potential to raise awareness of the issue, empower consumers to reduce their food waste and help Greece achieve its sustainability goals.*

Over the past few years, the food loss and waste (FLW) phenomenon and its negative economic, environmental and social effects has been considered one of the most important sustainability issues to be addressed at the global level. As FAO highlights, approximately one-third of the edible parts of food produced for human consumption globally was lost across the supply chain, which means around 1.3 billion tons of food loss and waste per year [1]. FLW reduction has been included among the 17 sustainable development goals of the UN's 2030 agenda and specifically in target 12.3 that aims to: "halve per capita global FW [food waste] at the retail and consumer levels and reduce food losses along production and supply chains" by 2030 [L1].

The purpose of the research was to quantitatively record the food discarded by households in Greece through the use of a mobile phone app. This marks the first time in Greece that FW is collected and recorded electronically. The app was tested for a period of two months – from late December 2021 to late February 2022 – with a small number of households to evaluate its functionality and identify potential issues, omissions and other observations. After the necessary improvements, the app's usage was expanded to a broader audience in early March 2022. This stage can be considered as a "pilot" phase, with the rationale that the app, being introduced for the first time in Greece, should first operate with a relatively small number of users before being used by all consumers and households in Greece.

The app was distributed to 125 user-households, aiming for the best representation of households with varying member counts. Our approach can be characterised as purposive or judgemental sampling. We deliberately chose the sample to ensure that participants could best serve the purposes and questions of our research. Thus, one of the main criteria used was the number of members in a household, as well as the household's monthly income and the region of residence. The usage of the app by households was voluntary, and users could